\documentclass[
   ,final            
numberedheadings 
  ]
  {aipproc}

\layoutstyle{6x9}

\begin{document}

\title{Vacuum Polarization Effects in the Worldline Variational Approach 
to Quantum Field Theory}



\author{R. Rosenfelder}{address={Particle Theory Group, Paul Scherrer 
Institute, CH-5232 Villigen PSI, Switzerland}
}

\begin{abstract}
The worldline variational approach 
is extended beyond the quenched approximation,
i.e. to include virtual pair production of heavy particles. This is achieved
either by an expansion of the functional determinant to second order or by an
hybrid {\it ansatz} for the quadratic trial action consisting of fields 
for the light particles and worldlines for the heavy ones 
as in the linear polaron model. Numerical results and analytic approximations 
show a reduction of radiative effects with increasing number of flavors.
\end{abstract}

\maketitle

\section{Introduction}
Worldline methods in Quantum Field Theory have seen a revival in 
recent years as an efficient technique in perturbative calculations \cite{Schub}
but also as a non-perturbative 
method to deal with strongly-interacting systems. We have extended 
the variational approach used by Feynman \cite{Fey} for the polaron problem 
to simple super-renormalizable scalar field theories \cite{WC} and 
to a renormalizable abelian gauge theory, viz. Quantum Electrodynamics \cite{QED}. 
In this approach fields are eliminated and replaced by (bosonic and fermionic) 
worldlines parametrized by the proper time. Recent progress
includes more general trial actions \cite{aniso}, 
Abraham-Lorentz-like equations for
the electron \cite{VALE} and treatment of the relativistic 
bound-state problem \cite{BRS}.

\section{Beyond the quenched approximation: expansion}
In all these applications the ``quenched approximation'' was made, i.e.  
vacuum polarization graphs have been neglected. This allowed the functional
integral over exchange fields (mesons, photons) to be performed.
Here we report 
on an attempt to include (part of) the vacuum polarization effects while
keeping the huge reduction in the number of degrees of freedom 
which is characteristic for the worldline
representation and  essential for a successful variational approximation.
In a simple scalar model where the heavy particles 
(''nucleons'') with mass  $M$ are described 
by an $N$-component field $ \Phi $ 
coupled to a ``meson'' field $\chi$ with mass $ \, m \>$ via 
$ \> {\cal L}_{\rm int} = g' \Phi^2 \chi \> $
this is achieved by expanding the functional determinant 
\begin{eqnarray}
\left ( \frac{\rm Det}{\rm Det_0}\right)^{-N/2} = {\rm Det}^{-\frac{N}{2}}  
\Bigl ( 1 + g' G_0 \chi \Bigr ) 
= \exp \Biggl \{-\frac{N}{2} {\rm Tr} 
\Bigl ( g' G_0 \chi-\frac{g'^2}{2} G_0 \chi  G_0 \chi
+ \ldots \Bigr ) \Biggr \} 
\end{eqnarray}
up to second order in the meson field. Here 
$ D_0 = (-\partial^2 - M^2)^{-1} $ is the {\it free} nucleon propagator.
After integration over 
$ \> \chi(x) \> $ the linear terms give rise to tadpoles which do not 
contribute to the dynamics 
whereas the quadratic terms change the interaction part of the 
$n$-particle worldline action to
\begin{eqnarray}
- \frac{g'^2}{2} \sum_{i,j=1}^n \, 
\int_0^{T_i} dt  \int_0^{T_j} dt' \int \frac{d^4 p}{(2 \pi)^4} \> 
\frac{\exp \left [ \, - i p \cdot \left (  x_i(t) -  x_j(t') \right ) \, 
\right ]}{p^2 - m^2 - \Pi_r (p^2) + i0} \> . 
\label{modified wl int}
\end{eqnarray}
Here 
\begin{eqnarray}
\Pi_r (p^2) = \frac{i}{2} g'^2 N \, \int \frac{d^4 k}{(2 \pi)^4} \, 
\frac{1}{k^2 - M^2 + i0} \, \frac{1}{(k-p)^2 - M^2 + i0} \, - \Bigl ( p^2 = m^2 \Bigr)
\end{eqnarray}
is the renormalized one-loop vacuum-polarisation contribution to the 
meson propagator. It can be easily shown that $\Pi_r(p^2)$ grows logarithmically 
at large $p^2$ so that the nucleon mass renormalization remains unchanged 
while at small $p^2$  we obtain a quadratic behaviour. This implies that 
the vacuum polarization effects
approximately amount to using an effective, reduced coupling constant
$ \> \alpha \to \alpha^{\star} = \frac{\alpha}{1 + N  \alpha/(6 \pi)} \> $.
Hence, in this approach vacuum-polarization insertions into all meson lines
are included (see Fig. 1a).
\begin{figure}
\vspace{-0.2cm}\hspace{2cm} 
\includegraphics[height=.2\textheight]{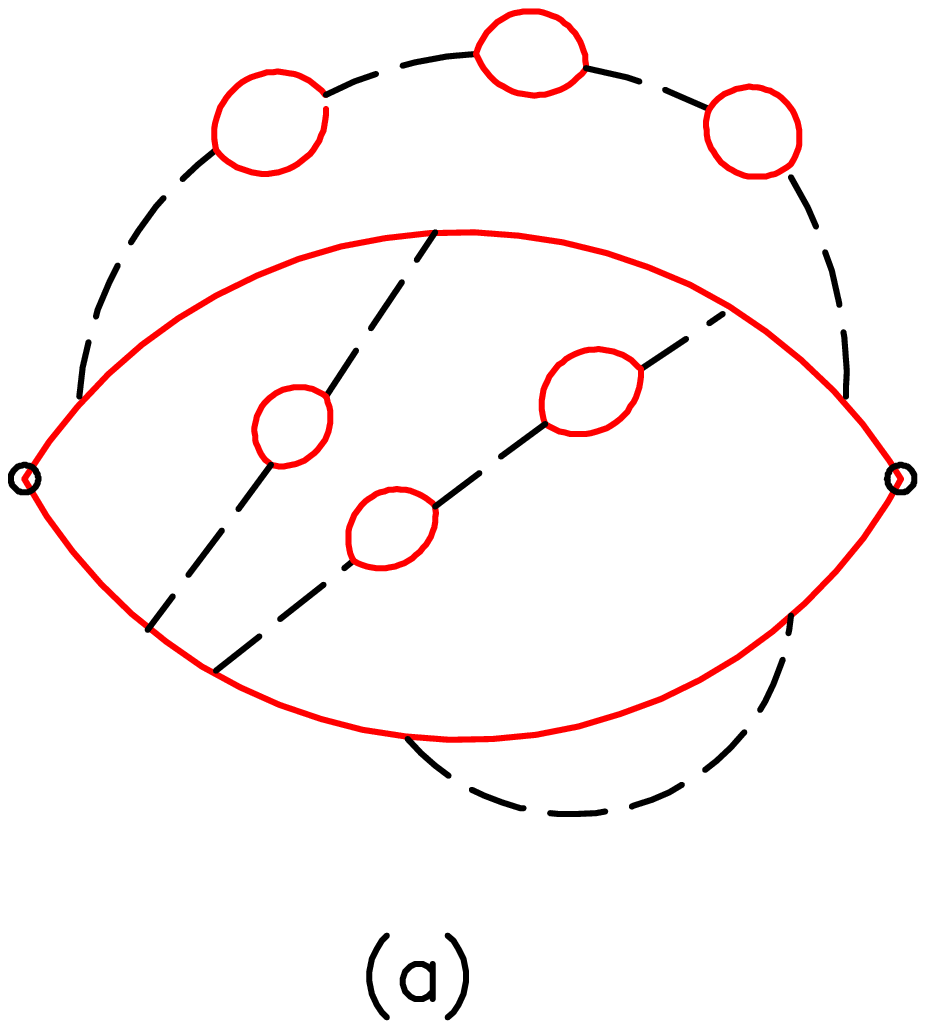} \hspace{2cm}
\vspace{0.2cm} \includegraphics[height=.2\textheight]{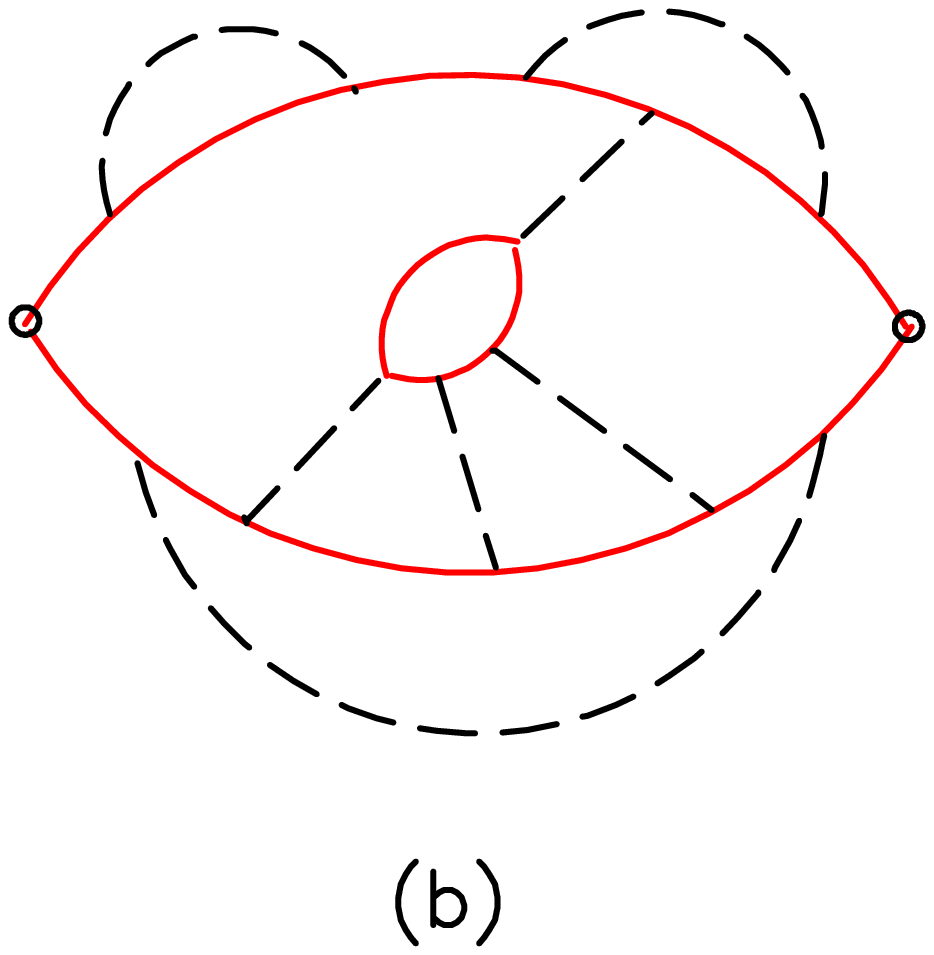}

\vspace*{0.5cm}

\caption{(a) Vacuum polarization insertions in the meson propagator 
(dashed line) for the
correlator whose pole determines the two-particle bound state energy
\cite{BRS}, (b) additional interactions of the virtually produced particle.}
\end{figure}
To see its effects we have re-calculated the meson-nucleon form factor at $q^2 = 0$ 
which experiences an enhancement due to radiative effects (as does the anomalous 
magnetic moment of the electron in QED) for various numbers of flavours $N$.
As shown in eq. (38) of ref. \cite{aniso} one of the variational parameters 
in the one-particle sector directly
determines the effective, physical coupling so that a numerical solution of the 
variational equations with the modified interaction (\ref{modified wl int})
gives the results shown in Fig. 2. As expected from the analytical approximation 
we observe a continous reduction  with $N$.

\begin{figure}
\hspace{0.5cm}
\includegraphics[height=.2\textheight]{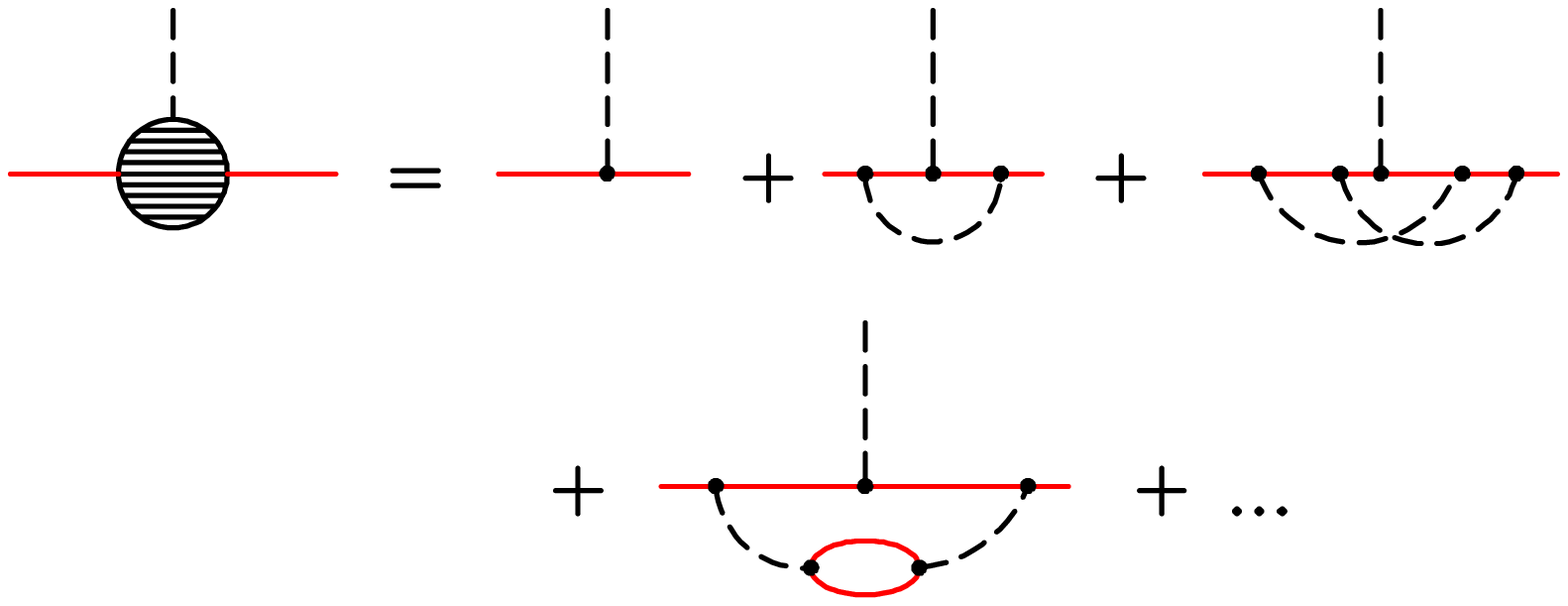}  
\hspace{1cm} \includegraphics[height=.3\textheight]{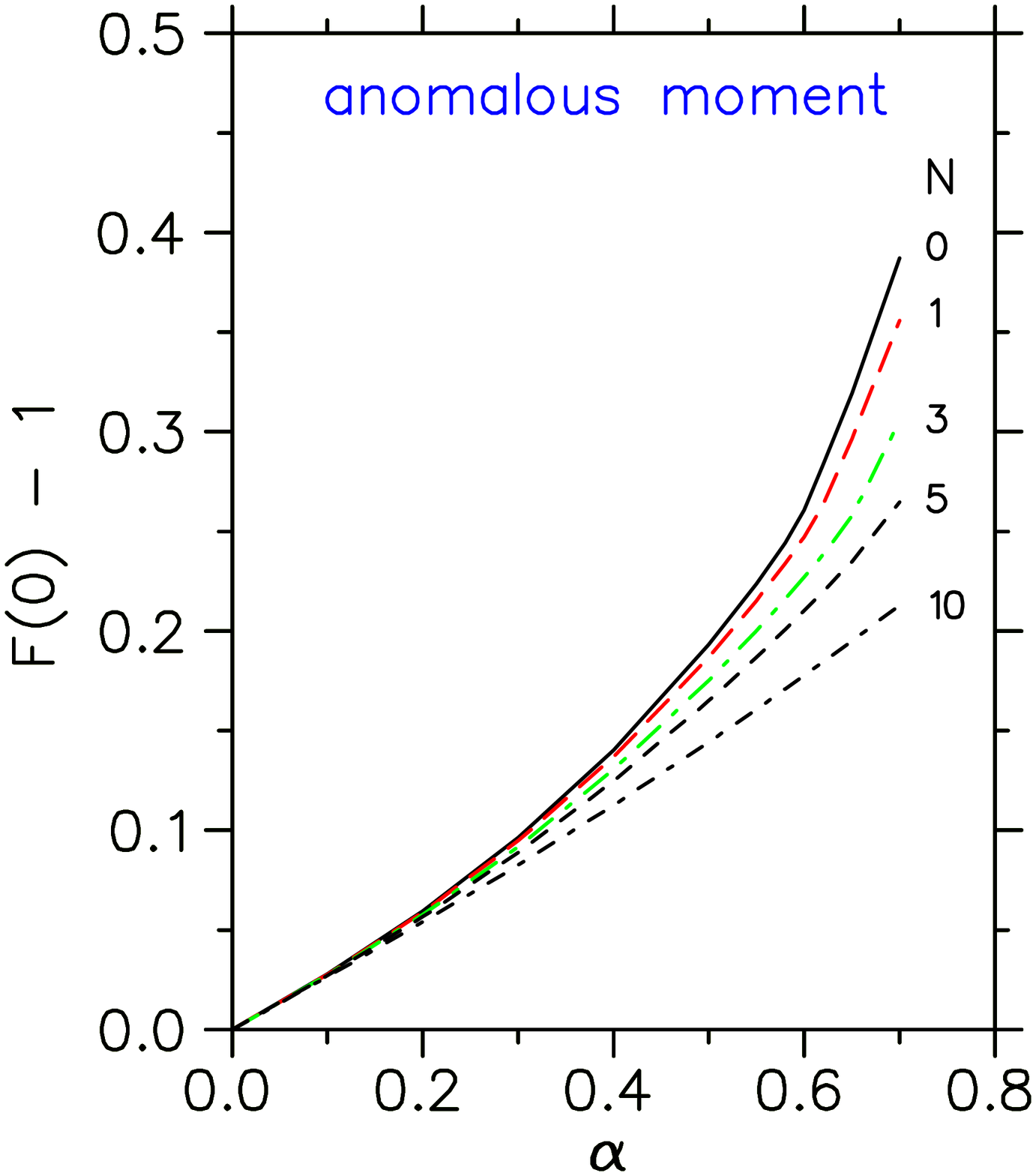} 

\caption{The anomalous moment in a perturbative expansion and non-perturbatively
as function of the dimensionless coupling constant 
$\alpha = g'^2/(16 \pi M^2) $ for different number of flavors $N$
when vacuum polarization effects are included.}
\end{figure}

\section{Beyond the quenched approximation: linear model}
The expansion of the determinant up to second order 
basically is a perturbative method and leaves
out interactions of the pair-produced heavy particles with
other nucleons, i.e. diagrams as shown in Fig.1b . To include these interactions 
in a non-perturbative way one has to find a (solvable!) trial action which 
contains meson fields and worldlines for the nucleons at the same time. This is
provided by the linear polaron model of Bogoliubov \cite{Bog} which is motivated
by replacing the non-linear coupling between the $\chi$-field and each nucleon 
worldline $x_i(t)$ by a linear one:
\begin{eqnarray}
g' \, \chi(x_i(t)) =   g' \, \int \frac{d^4 k}{(2 \pi)^2} \> 
\underbrace{e^{i k \cdot x_i(t)}}_{\longrightarrow \>  
\ell_0(k^2) + \ell_1(k^2) \, i k \cdot x_i(t)} \> \tilde \chi(k) \> .
\end{eqnarray}
In the new trial action $\ell_j(k^2)\, , \, j = 0, 1 $ and the unquenched meson 
propagator are taken as variational functions. By construction, interactions 
as depicted in Fig. 1b are now included and the averages needed for the Feynman-Jensen 
variational principle
can be worked out analytically. A detailed account will be given elsewhere.



\begin{thebibliography}{99}
\bibitem{Schub} C.~Schubert,
\emph{Phys. Rept.}  \textbf{355}, 73 -- 234 (2001).

\bibitem{Fey} R.~P.~Feynman, \emph{Phys. Rev.} \textbf{97}, 660 -- 665 (1955).

\bibitem{WC}
R.~Rosenfelder, and A.~W.~Schreiber,  \emph{Phys. Rev. D} \textbf{53}, 3337 --
3353,
3354 -- 3365 (1996);\\
A.~W.~Schreiber, R.~Rosenfelder, and C.~Alexandrou, 
\emph{Int. J. Mod. Phys. E} \textbf {5}, 681 -- 716 (1996);\\
A.~W.~Schreiber, and R.~Rosenfelder, 
\emph{Nucl. Phys. A} \textbf {601}, 397 -- 424 (1996); \\ 
C.~Alexandrou, R.~Rosenfelder, and A.~W.~Schreiber,
\emph{Nucl. Phys. A} \textbf {628}, 427 -- 457 (1998); \\
N.~Fettes, and R. Rosenfelder, \emph{Few-Body Syst.} \textbf {24}, 1 -- 25 (1998).

\bibitem{QED}
C.~Alexandrou, R.~Rosenfelder, and A.~W.~Schreiber, \emph{Phys. Rev. D} 
\textbf{62}, 085009 (2000).

\bibitem{aniso}
R.~Rosenfelder, and A.~W.~Schreiber, \emph{Eur. Phys. J. C} \textbf{25}, 139 -- 
156 (2002).

\bibitem{VALE}
R.~Rosenfelder, and A.~W.~Schreiber, \emph{Eur. Phys. J. C} \textbf{37}, 161 -- 
172 (2004).

\bibitem{BRS}
K.~Barro-Bergfl\"odt, R.~Rosenfelder, and M.~Stingl, \emph{Mod. Phys. Lett. A}
\textbf{20},  2533 -- 2543 (2005).

\bibitem{Bog} 
N.~N.~Bogoliubov, and N.~N.~Bogoliubov, Jr.,  \emph{Sov. J. Part. Nucl.} 
\textbf{11}, 93 --110 (1980). 












 
\end{thebibliography}
\end{document}